\documentclass[]{interact}

\usepackage{epstopdf}
\usepackage[caption=false]{subfig}

\usepackage[numbers,sort&compress,merge]{natbib}
\bibpunct[, ]{[}{]}{,}{n}{,}{,}
\renewcommand\bibfont{\fontsize{10}{12}\selectfont}

\theoremstyle{plain}

\theoremstyle{definition}

\theoremstyle{remark}

\title{Spectroscopic properties of the molecular ions BeX$^+$ (X=Na, K, Rb): Forming cold molecular ions from an ion-atom mixture by stimulated Raman adiabatic process}

\author{
\name{Hela Ladjimi\textsuperscript{a}, Dibyendu Sardar\textsuperscript{b}, Mohamed Farjallah\textsuperscript{a}, Nisrin Alharzali\textsuperscript{a}, Somnath Naskar\textsuperscript{b,e}, Rym Mlika\textsuperscript{a}, Hamid Berriche\textsuperscript{a,c} and Bimalendu Deb\textsuperscript{b,d}}
\affil{\textsuperscript{a}Laboratory of Interfaces and Advanced Materials, Faculty of Sciences of Monastir, University of Monastir, 5019 Monastir, Tunisia; \textsuperscript{b}Department of Materials Science, Indian Association for the Cultivation of Science (IACS), Jadavpur, Kolkata 700032, INDIA; \textsuperscript{c}Department of Mathematics and Natural Sciences, School of Arts and Sciences, American
University of Ras Al Khaimah, RAK, P.O. Box 10021, UAE $^*$Corresponding author; \textsuperscript{d}Raman Center for Atomic, Molecular and Optical Sciences, Indian Association for the
Cultivation of Science, Jadavpur, Kolkata 700032, India; \textsuperscript{e}Department of Physics, Jogesh Chandra Chaudhuri College, Kolkata 700033, INDIA.}
}
\begin{document}
\maketitle

\begin{abstract}
In this theoretical work, we calculate potential energy curves, spectroscopic parameters and transition dipole moments of molecular ions BeX$^+$ (X=Na, K, Rb) composed of alkaline ion Be and alkali atom X with a quantum chemistry approach based on the pseudopotential model, Gaussian basis sets, effective core polarization potentials, and full configuration interaction (CI). We study in detail collisions of the alkaline ion and alkali atom in quantum regime. Besides, we study the possibility of the formation of molecular ions from the ion-atom colliding systems by stimulated Raman adiabatic process   and discuss the parameters regime under which the population transfer is feasible. Our results are important for ion-atom cold collisions and experimental realization of cold molecular ion formation.
\end{abstract}

\begin{keywords}
Pseudo potentials; Configuration Interaction; Spectroscopic constants; elastic scattering; dark state; STIRAP; molecular ion
\end{keywords}

\section{Introduction}

In recent years, cold molecular ion  \cite{carr} have received considerable attention in
respect of cooling and trapping of molecules \cite{hummon} and
high-resolution spectroscopy \cite{stwalley}. Because of the characteristic long range ion-atom
potentials, cold ion-atom collisions \cite{cote,idziaszek,krych} are different from atom-atom and ion-ion
collisions. The long range ion-atom potential described by -$C_4/r^4$ , where $C_4 = \alpha q^2/ (8\pi\epsilon_0)$,
with $\alpha$ being the static electric polarizability of the atom and $r$ the ion-atom separation. The
scattering in this potential offers the primary channel for the exchange of energy
between the ions and atoms. Accordingly, the investigations of ion-atom collisions at low
energy are very useful for the study of sympathetic cooling of atomic and ionic translational motion
\cite{ravi,sivarajah} and to get a detailed control of the internal and external degrees of freedom of the molecular ions \cite{boyarkin}. Ion-atom collisions at low energy are important for understanding charge transport phenomena \cite{charge}, ion-atom bound states \cite{bound-states} and cold molecular ions \cite{cold-ion,cold-ion1}. Also, a wealth of cold molecular ionic species could be created, opening the way to a rich chemistry at temperatures of a few mK, or less \cite{willitsch}. Besides, controlled ion-atom cold collisions may be used for future quantum information processing \cite{information,information1}. Again, ion-atom collisions are involved with reactive charge exchange process \cite{charge-transfer,charge-transfer1}. In a recent experimental work, it is shown that the internal electronic states of a single trapped ion can be manipulated via spin exchange processes \cite{kohl}. Furthermore, the role of the hyperfine interaction on a chemical reaction of ion-atom colliding system is studied \cite{Ratschbacher}.

Many experimental groups have carried out experiments with
various combinations of alkali atoms Rb and alkaline-earth ions: Rb atoms with Ca$^+$
\cite{hall,hall1} and Ba$^+$ \cite{hall2} ions. In addition, an optical dipole trap of Rb atoms has been merged in a Paul trap containing a few Ba$^+$ ions \cite{schmid,schmid1}. Radiative emission during cold collisions between trapped laser-cooled Rb atoms and alkaline-earth ions Ca$^+$, Sr$^+$ and Ba$^+$, and various spectroscopic properties of the related molecular ions are studied theoretically by Aymar $et$ $al$ \cite{silva}. Theoretically, first ab initio calculations of alkaline ion and hydrogen atom of (BeH)$^+$ molecular ion system have been
carried out by Ornellas $et$ $al$ \cite{ornellas,orenellas1,orenellas2,orenellas3}. The molecular ion (BeLi)$^+$ is extensively investigated using only the ground and the first excited states \cite{alexander}. The 
low-energy collisions for the alkali ion and an alkali atom system have also been performed \cite{rakshit1}.  Recently, Rakshit $et$ $al$ \cite{rakshit}
have theoretically investigated (BeLi)$^+$ system at low energy predicting
the possibility of formation of the cold molecular ion by photoassociation (PA) process.

A promising way to form cold molecule or molecular ion is two-photon Raman process or Stimulated Raman adiabatic process (STIRAP) \cite{kral2007,cbb_bbc_bergmann}. STIRAP is an important population transfer method occurred in a coherent way by which population of the initially populated level can be efficiently transferred to the final target level. To make effective state transfer by this process, two lasers are applied in a counter intuitive way so that the system will remain in a dark state (DS) satisfying an adiabatic condition.  This method was first implemented in $\Lambda$-type discrete three-level systems \cite{kuklinski1989} where
DS, a special eigen state of the system of which population remains zero during transfer process. In these discrete $\Lambda$ systems complete population transfer is possible. But systems involving a continuum such as two-atom or ion-atom collisional continuum, complete population transfer is difficult by this coherent method as it is impossible to form a transparent DS at all energies. In a recent theoretical paper on continuum-bound-bound system resembling $\Lambda$-type configuration \cite{dibyendu}, it is shown that the population in the continuum of ion-atom scattering states can be partially transferred to a target state via STIRAP-like fashion and the system follows a quasi-DS condition. 
\begin{figure}
\begin{center}
 \includegraphics[width=0.73\linewidth]{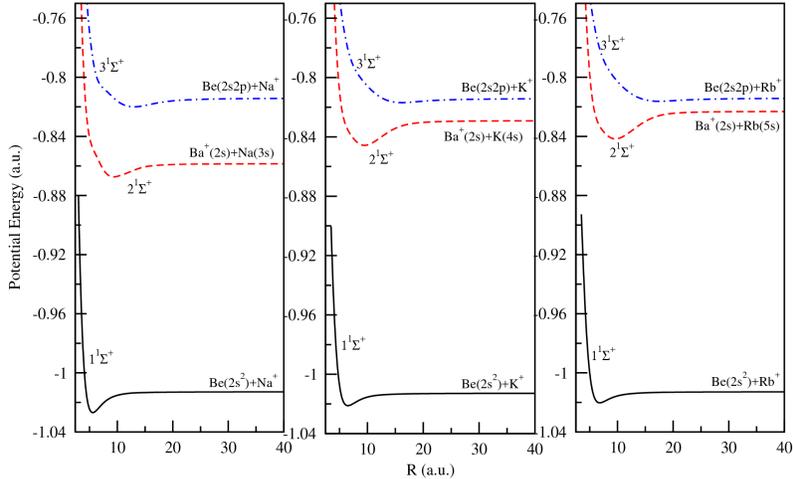}
 \caption {Potential energy curves of three BeX$^+$ systems are plotted as a function of internuclear separation in different electronic states, respectively. In the lowest panel, the curves are plotted in ground state where solid black line corresponds to (BeNa)$^+$ system, red dashed and blue double dashed lines are for (BeK)$^+$ and (BeRb)$^+$, respectively. Next two panels represent the same for the electronic states 2$^1\Sigma^+$ and 3$^1\Sigma^+$ of three systems.   }
 \label{potential}
 \end{center}
\end{figure}
 
This study has three parts. In the Part-I, 
 we calculate adiabatic potential energy curves for (BeNa)$^+$, (BeK)$^+$ and (BeRb)$^+$ molecular ionic systems by ab initio approach involving a non-empirical pseudo-potential for the X$^+$(X=Na, K, Rb) and Be$^{2+}$ cores. Here, core-valence correlation is accounted in an empirical form of core-polarization potentials and multireference configuration interaction technique is employed. In the potential energy plot, short range part is obtained by our ab initio calculation and long-range potential obtained by the sum of the dispersion terms described by -1/2($C_4/r^4+C_6/r^6$). We calculate spectroscopic constants and transition dipole moment for these systems. In Part-II of this work, 
we investigate elastic hetero nuclear cold collisions for a wide range of energies ranging from 0.01
 $\mu$K to 1 K between alkaline-earth ion Be$^+$ and neutral alkali atom X (X=Na, K, Rb) of each BeX$^+$ system in their first excited electronic state. At low energy regime, hetero-nuclear ion-atom collision is dominated by elastic scattering process and charge transfer process is suppressed. 
 
 In Part-III of this study, we explore formation of cold molecular ion from a colliding pair of alkaline ion Be$^+$ and alkali atom X by photoassociative Raman process. Following the Fano diagonalization method \cite{fano}, we obtain a dressed continuum state and this state can be treated as a quasi-DS under certain physical conditions. We apply a STIRAP-like process for transfer of the atomic population from ion-atom scattering continuum of first excited state to molecular population in electronic ground state. We have found that effective population transfer is possible for three BeX$^+$ systems.

The plan of this paper is as follows. In Part-I, we briefly describe the ab initio method
used in the calculation of the Born-Oppenheimer potential energy curves. Numerical results of
the electronic structure namely interaction potential and transition dipole moments of these
molecular ion systems are presented and discussed in this portion. In Part-II, we study
the elastic scattering properties between the alkaline-earth ion Be$^+$ and the alkali
atom X (X=Na, K, Rb) using the first excited electronic state. The formation of cold molecular ions by STIRAP technique are
presented and discussed in Part-III. In last section we make some remarkable conclusions.
\section{Part-I: Calculation of molecular and spectroscopic properties}
\subsection{Method}

As in previous works on LiH \cite{ab-in}, LiNa \cite{mabrouk} and LiCs \cite{mabrouk1}, the BeX$^+$
(X=H and Li) molecular ion \cite{farjallah1,farjallah2} is treated as a two-valence electron
system using the non-empirical pseudopotential of Barthelat and Durand
\cite{barthelat}, in its semi-local form and used in many previous works \cite{ab-in,mabrouk,mabrouk1,berriche,berriche1,ghanmi,ghanmi1,ghanmi2,bouzouita,ghanmi3,berriche2}. Furthermore, the self-consistent field calculation (SCF) is followed by a
full valence configuration interaction (FCI) calculation using the CIPCI
algorithm of the standard chain of programs of the Laboratoire de Physique Quantique of Toulouse. For the simulation of the interaction
between the polarizable Be$^{2+}$ core with the valence electrons of the alkali
nucleus, a core polarization potential $V_{CPP}$ is used, according to the
operator formulation of M\"{u}ller, Flesh and Meyer \cite{muller}:
\begin{equation}
 V_{CCP}=-1/2 \sum_\lambda \alpha_\lambda \vec f_\lambda.\vec f_\lambda
\end{equation}
where $\alpha_\lambda$ is the dipole polarizability of the core $\lambda$
and $\vec f_\lambda$ is the electric
field produced by valence electrons and all other cores on the core $\lambda$.
\begin{equation}
 f_\lambda=\sum_i \frac{r_{i\lambda}}{r^3_{i\lambda}}F(\vec r_{i\lambda},\rho_\lambda)-\sum_{\lambda'\ne\lambda}\frac{R_{\lambda'\lambda}}{R^3_{\lambda'\lambda}}Z_\lambda
\end{equation}
where $\vec r_{i\lambda}$ is a core-electron vector and $\vec R_{\lambda'\lambda}$ is a core-core vector. According to the formulation of Foucrault $et$ $al$ \cite{foucrault} the cut-off function $F(\vec r_{i\lambda},\rho_\lambda)$ 
is taken to be a function of $l$ in order to treat separately the
interaction of valence electrons of different spatial symmetry and the core
electrons. It has a physical meaning of excluding the valence electrons
from the core region for calculating the electric field. 
\begin{table}[ht]
 \caption{Polarizabilities, cut-off radii and pseudo-potential parameter (in a.u.) atoms.}
  \centering  
  \begin{tabular}{c c c c c c }
   \hline\hline
   
     Atom  &      $\alpha$  &       $\rho_s$  &       $\rho_p$  &   $\rho_d$  \\ [0.5ex]
    \hline
         
     Be(Z=4)  &      0.0519      &     0.889    &         0.882  &     1.14         \\
         
     Na(Z=11)   &   0.9930        &      1.4423   &        1.625  &    1.500         \\
         
     K(Z=19) &     5.3450         &       2.0670   &       1.9050  &   1.500        \\ 
     
     Rb(Z=37) &    9.245          &       2.5213  &         2.279  &   2.511          \\ [1ex]

   \hline     
  \end{tabular}
  \label{table1}
\end{table}

For the Beryllium, Sodium, Potassium and Rubidium atoms, we used
(7s9p10d/7s8p9d) \cite{farjallah1,farjallah2}, (7s6p5d3f/6s5p4d2f) \cite{berriche3}, (7s5p7d1f/6s5p5d1f)
\cite{magnier} and (7s4p5d1f/6s4p4d1f) \cite{Pavolini},
basis set of Gaussian-type orbital,
where diffuse orbital exponents have been optimized to reproduce the
atomic states 2s, 2p, 3s, 3p and 3d, and 2s$^2$, 2s2p, 2s3s, 2p$^2$, 2s3p and
2s3d for Be$^+$ and Be species, respectively. Following the formulation of
Foucrault $et$ $al$ . \cite{foucrault}, cut-off functions with $l$-dependent adjustable
parameters are fitted to reproduce not only the first experimental
ionization potential but also the lowest excited states of each $l$ for Na, K
Rb, Be$^+$ and Be species. In the present work, the core polarizability of Be$^{2+}$
s taken to be $\alpha_{Be^{2+}}=$0.0519 $a^3_0$ \cite{muller}, 0.993 $a^3_0$, 5.354 $a^3_0$  and
9.245 $a^3_0$ the optimized cut-off parameters for the lowest valence s, p
and d one-electron states of the Na, K, Rb and Be atom are represented
respectively in table \ref{table1}.

\begin{table}[ht]
 \caption{Asymptotic energy of the alkali earth BeX$^+$(X=Na, K, Rb) of X$^1\Sigma^+$, 2$^1\Sigma^+$, 3$^1\Sigma^+$ electronic states in cm$^{-1}$.}
  \centering  
  \begin{tabular}{c c c c c c }
   \hline\hline
   
     State          &      Asymptotic limit  &    Our work  &      Experimental \cite{new1} & $\mid \Delta E\mid$ \\ [0.5ex]
    \hline
         
     X$^1\Sigma^+$  & Be(2s$^2$)+Na$^+$     & -222590.66     &      -222075.44               &  515.22             \\
     
      2$^1\Sigma^+$  & Be$^+$(2s)+Na(3s)    & -188337.2     &      -188331.0                 &    6.2           \\
     
     3$^1\Sigma^+$  & Be(2s2p)+Na$^+$    & -178998.76    &      -179510.09                &    511.33          \\
     
     \\
         
     X$^1\Sigma^+$  & Be(2s$^2$)+K$^+$   & -222590.66     &      -222075.44               &    515.22         \\
     
     2$^1\Sigma^+$  & Be$^+$(2s)+K(4s)    & -181897.14     &      -181891.15                &    5.99       \\
     
     3$^1\Sigma^+$  & Be(2s2p)+K$^+$    & -178998.76    &      -179510.09                &    511.33          \\
     
     \\
     
     X$^1\Sigma^+$  & Be(2s$^2$)+Rb$^+$     & -222590.66     &      -222075.44               &  515.22             \\
     
      2$^1\Sigma^+$  & Be$^+$(2s)+Rb(5s)    & -180578.32     &      -180572.17                 &    6.14           \\
      
      3$^1\Sigma^+$  & Be(2s2p)+Rb$^+$    & -178998.76    &      -179510.09                &    511.33          \\
                   
   \hline     
  \end{tabular}
  \label{table-new}
\end{table}

\subsection{Electronic structure of BeX$^+$ (X=Na, K, Rb)}

The best way to check the accuracy of the used basis sets is to calculate the atomic energy levels and then the molecular asymptotic limits. For this aim, the molecular asymptotic energies were determined for Be(2s$^2$, 2s2p, 2s3s, 2p$^2$, 2s3p and 2s3d) + X$^+$(X=Na, K and Rb) and Be$^+$ (2s, 2p, 3s3p and 3d)+ X, for which Na(3s, 3p) and K (4s, 4p) nad Rb(5s, 5p) are involved. The comparison with experimental limits shows a very good agreement. The molecular asymptotic energies for the electronic states (X-2-3)$^1\Sigma^+$ states are illustrated in table \ref{table-new}. They are compared to the experimental limits \cite{new1}. From this table, we found that the difference betwen our limits and experimental values does not exceed 515.22 cm$^{-1}$, which is found for Be (2s$^2$)($^2$S) atomic levels corresponding to a relative difference ($\Delta E/E$) of 0.23$\%$. The quality of the used basis sets and cutoff radii optimized for the pseudopotentials are confirmed by this good agreement. In fact, the theoretical limits for the higher excited states are also in very good agreement with experimental ones. These asymptotic energies will be present in a separate paper devoted to ab initio calculation for many electronic states of $^{1,3}\Sigma^+$ and $^{1,3}\Pi$ and symmetries, permanent and transition dipole moments, vibrational analysis, and relative lifetimes for the three ionic molecular systems BeNa$^+$, BeK$^+$ and BeRb$^+$.

The BeX$^+$ (X=Na, K, Rb) systems are modeled as molecules with two valence
electrons moving in the field of the Be$^{2+}$ and X$^+$ (X=Na, K, Rb). A full configuration
interaction method is used to obtain the relevant potential energy curves. The potential
energy for the molecular ionic systems (BeNa)$^+$, (BeK)$^+$ and (BeRb)$^+$ are
calculated for a large and dense grid of internuclear distances ranging
from 4 to 200 a.u. Limited to the X, A and C state, the potential energy curves for each
molecule (BeNa)$^+$, (BeK)$^+$ and (BeRb)$^+$ plotted in figure \ref{potential}. These states dissociate
 into Be(2s$^2$)+X$^+$, Be$^+$(2s)+X(ns), Be(2s2p)+X$^+$ for BeX$^+$ systems.
  Similar to the (BeH)$^+$ and (BeLi)$^+$ ionic systems \cite{farjallah1,farjallah2}, the molecular ground
state is correlated at large distance to the neutral alkaline-earth atom, Be in its ground level,
and a closed-shell alkali-metal ions X$^+$. An accurate analysis of these curves show
that the ground state has the deepest well compared to the 2$^1\Sigma^+$ and 3$^1\Sigma^+$
excited states.

\begin{figure}
\begin{center}
 \includegraphics[width=0.6\linewidth]{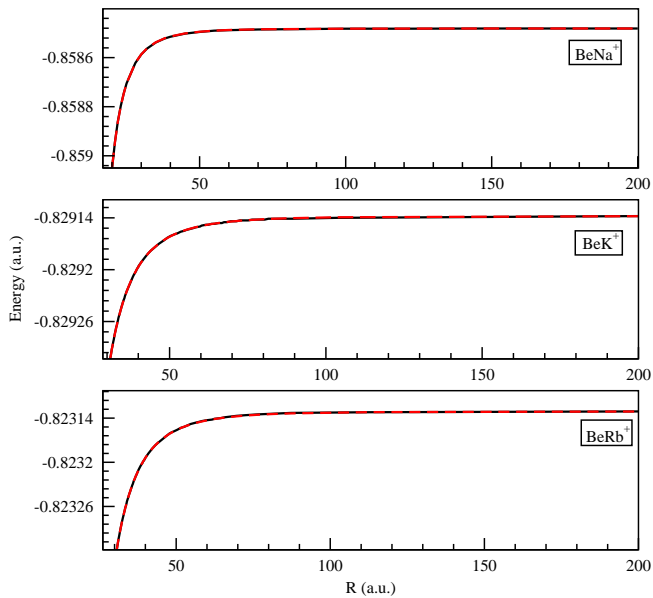}
 \caption {Comparison between our ab initio energies (black line) and analytical
 energies (red line) of the first excited electronic state (2$^1\Sigma^+$) of BeX$^+$ systems at long-range.}

 \label{fit}
 \end{center}
\end{figure}

Using the interpolation of the vibrational levels by least-squares approach, the
spectroscopic constant (the equilibrium distance $R_e$, well depth $D_e$, the frequency of 
vibration $\omega_e$, frequency and anharmonicity $\omega_e \chi_e$ and rotational constant B$_e$ ) for the three electronic 
states of each molecular ionic systems were calculated and presented in table \ref{table2-new}. It is clear from
figure \ref{potential} and table \ref{table2-new} that the equilibrium positions of potential energy
curves in all considered states evidently increase from (BeNa)$^+$ to (BeRb)$^+$
molecular ionic systems. This phenomenon is due to the increase in
electronic shells (atomic radii) from Na to Rb atoms.

\begin{table}[ht]
 \caption{Spectroscopic constants of the ground and two excited singlet electric states of (BeNa)$^+$, (BeK)$^+$ and (BeRb)$^+$ molecular ion systems, respectively.}
  \hspace{-1.350in} 
  \begin{tabular}{c c c c c c c c c c c}
   \hline\hline
   
 & State  &  R$_e$(a.u.) &   D$_e$ (cm$^{-1}$) &  T$_e$ (cm$^{-1}$)  &$\omega_e$ (cm$^{-1}$)  & $\omega_e x_e$ (cm$^{-1}$)&  B$_e$ (cm$^{-1}$)& Methods/Basis& Ref\\ [0.5ex]
    \hline
 &    BeNa$^+$ &         &                     &                      &                        &                           &                   &               &   \\
    
 & X$^1\Sigma^+$ & 5.58  &   3079              &     0                &          200.33        &   3.25                    & 0.299203          &   CIPSI        &This work\\ 
 
  &              & 5.73  &   2764              &     0                &          182.75        &   3.02                    & 0.283423       & CCSD(T)/AQZP       &This work\\
  
 &              & 5.75  &   3027              &     0                &          188.24        &   2.93                    & 0.281377       & MP2/AQZP       &This work\\
 
 &              & 5.686  &            &     0                &          202       &                    &       & MP2/6-31G$^*$       & \cite{pykk}\\
 
  &              & 5.731  &            &     0                &          199       &                    &       & HF/6-31G$^*$       & \cite{pykk}\\
  
 &              & 5.624  &   3064.7     &     0                &     199.0        &   3.2               &        & CCSD(T)/cc-pCVQZ       &\cite{fedrov}\\ 
 
 &              & 5.617  &   3018.8     &     0                &     193.0        &   1.9               &        & MRCI/cc-pCVQZ       &\cite{fedrov}\\ 
 
 &              & 5.614  &   3085.2     &     0                &     194.6        &   1.9               &        & MRCI/aug-cc-pCV5Z       &\cite{fedrov}\\ 
 
 & 2$^1\Sigma^+$ & 9.33  &  1961           &     34998      &          93.06        &  1.48             &    0.107130          &   CIPSI        &This work\\
 
 & 3$^1\Sigma^+$ & 12.92  &  1300          &     45378     &          53.00       &  0.54             &    0.055720         &   CIPSI        &This work\\
 
 \\
 
 &    BeK$^+$ &         &                     &                      &                        &                           &                   &                &   \\
 
 & X$^1\Sigma^+$ & 6.52 &   1833          &     0                &          135.61        &   2.96                    & 0.193599          &   CIPSI        &This work\\
 
 &               & 6.74 &   1700          &     0                &          140.63        &   2.91                    & 0.181005          &   CCSD(T)       &This work\\ 
 
 & 2$^1\Sigma^+$ & 9.48  &  3638           &     38515      &          88.06        &  0.53             &    0.091501          &   CIPSI        &This work\\
 
 & 3$^1\Sigma^+$ & 16.12  &  593          &     44838     &          33.52      &  0.47             &    0.031589         &   CIPSI        &This work\\ 
 
 \\ 
 
&    BeRb$^+$ &         &                     &                      &                        &                           &                   &                &   \\
    
& X$^1\Sigma^+$ & 6.81  &   1630              &     0                &     119.74       &   2.50                    & 0.159460          &   CIPSI        &This work\\ 
 
&              & 7.19  &  1399           &     0                &          112.35        &   2.25                   & 0.176324      & CCSD(T)/AQZP       &This work\\ 

&              & 7.14  &  1505           &     0                &          106.54        &   1.88                   & 0.145486      & MP2/AQZP       &This work\\

 & 2$^1\Sigma^+$ & 9.61  &  4072           &     39198      &          80.53        &  0.39             &    0.079968          &   CIPSI        &This work\\
 
 & 3$^1\Sigma^+$ & 17.07  &  466          &     44764     &          25.98      &  0.36            &    0.025346         &   CIPSI        &This work\\

   \hline     
  \end{tabular}
  \label{table2-new}
\end{table}

The long-range potential interaction of ground states X$^1\Sigma^+$, for the three systems, varies as
-$C_4/r^4$ where $C_4$ is related to the static polarizability of neutral Beryllium  atom. On the other hand, the long rang part of 2$^1\Sigma^+$ states varies also as $r^{-4}$ , with a coefficient depending on the static polarizability of the alkaline neutral species. To check the accuracy of our calculated Potential Energy Curves
and in order to prove its behavior at long-range, we have interpolated the
potential energy curve of the ground state X$^1\Sigma^+$, dissociating into Be($^1S$)+X$^+$($^1S$),
and the first excited state 2$^1\Sigma^+$, dissociating into Be$^+$($^2S$)+X($^2S$).
Based on
the fact that the long range ion-atom potential interaction described by -$C_4/r^4$ and by
square fitting of the long-range part of our curves with a single parameter
give $C_4$ values 162.3 a.u, 294.11 a.u and 310.85 a.u  for Na($^2S$), K($^2S$) and Rb($^2S$), respectively. As expected, the comparison of our polarizability level with the
available theoretical and experimental results shows a very good
agreement. Derevianko $et$ $al$ \cite{derevianko}, Ekstrom $et$ $al$ \cite{ekstrom} and Holmgren $et$ $al$ \cite{holmgren}
found the $C_4$ values 162.6, 162.7($\pm$0.8) and 162.7($\pm$1.2) a.u, respectively for Na($^2S$).
Derevianko $et$ $al$ \cite{derevianko}, Lim $et$ $al$ \cite{lim} and Molof $et$ $al$ \cite{molof} and Holmgren $et$
$al$ \cite{holmgren} found $C_4$ values 289.1, 291.1, 293($\pm$6) and 290.6($\pm$1.1) a.u for
K($^2S$) and for Rb($^2S$), they obtained the values 318.6, 316.2, 316.2($\pm$6)
and 318($\pm$1.4). We reported in figure \ref{fit} a comparison between the ab initio
and analytical energies of the first excited states 2$^1\Sigma^+$  of BeX$^+$ molecular
ions. These figures show satisfactory agreement between the ab initio and
analytical energies.

No experimental data nor theoretical calculations have been reported yet for BeK$^+$ and BeRb$^+$ molecular ions. However, Pyykk\"{o} \cite{pykk} investigated theoretically, BeNa$^+$ in its ground state using the MP2 and HF level of calculations. Our results for BeNa$^+$ are in good concordance with those presented by Pyykk\"{o}. The spectroscopic constants provided by Pyykk\"{o} \cite{pykk} were calculated for the ground state X$^1\Sigma^+$, using two methods: M\o{}ller-Plesset MP2 (full) and HF methods with polarized split-valence basis set 6-31G$^*$. They found using both approaches (MP2 and HF), respectively, R$_e=5.686$ au and $\omega_e=202$ cm$^{-1}$ and R$_e=5.731$ and $\omega_e=199$ cm$^{-1}$. The spectroscopic constants given by the MP2 calculations are, as expected, closer to our values (R$_e=5.58$ au and $\omega_e=200.33$ cm$^{-1}$) than those found by the HF approximation. Very recently, Fedorov et al \cite{fedrov} presented the spectroscopic constants of the ground state of BeNa$^+$ using different methods and basis sets (see table \ref{table2-new}). We note that the spectroscopic constants obtained using the CCSDT level and cc-pCVQZ basis set (R$_e=5.624$ au , D$_e=3064.7$ cm$^{-1}$, $\omega_e=199.0$ cm$^{-1}$ and $\omega_e$ x$_e=3.2$ cm$^{-1}$) are in excellent agreement with our values. The spectroscopic constants with MRCI level of calculation using two different basis sets (cc-pCVQZ and aug-cc-pCV5Z) are similar to each other (R$_e=5.617$ au , D$_e=3018.8$ cm$^{-1}$, $\omega_e=193.0$ cm$^{-1}$ and $\omega_e$ x$_e=1.9$ cm$^{-1}$ using the cc-pCVQZ; and R$_e=5.614$ au , D$_e=3086$ cm$^{-1}$, $\omega_e=194.6$ cm$^{-1}$ and $\omega_e$ x$_e=1.9$ cm$^{-1}$ using the cc-pCV5Z). As it was mentioned by Fedorov et al \cite{fedrov}, the discrepancy of the harmonicity constant values between the two methods ($\omega_e$ x$_e=3.2$ cm$^{-1}$ in CCSDT level and $\omega_e$ x$_e=1.9$ cm$^{-1}$ in MRCI level) can be explained by the fact that the CCSDT method produces a more accurate potential energy curve near the equilibrium distance than the MRCI. The comparison of our spectroscopic constants with the recent molecular constants given by Fedorov et al \cite{fedrov} shows a very good agreement with relative differences $\Delta$(R$_e$)$<$0.8$\%$, $\Delta$(D$_e$)$<$1.96$\%$ and $\Delta$($\omega_e$)$<$3.7$\%$. This very good agreement confirms the quality and the accuracy of our CIPSI calculations and, therefore, the use of the pseudopotential technique.

To the best of our knowledge, no experimental data nor theoretical calculations have been reported yet for BeK$^+$ and BeRb$^+$ molecular ions. Therefore, their spectroscopic properties are reported here for the first time. Consequently, we think that it is worthwhile to perform other calculations for the ground state and some lower excited states for each molecular system. Table \ref{table2-new} shows the Spectroscopic constants of BeX$^+$ (X=Na, K Rb) ionic systems using the quantum chemistry package Molpro \cite{molpro} and using different basis sets and approaches. In this context, CCSD(T) and MP2 calculations were carried out for BeNa$^+$ using AQZP \cite{fantin} basis set for both Be and Na atoms. The agreement between the CCSD(T) and MP2 approaches and the CIPSI calculations for the ground state is remarkable, which confirms the high quality of our calculations and validate again the use of the pseudopotential technique to reduce time and to use large basis sets optimized to get accurate highly excited states. The comparisons between our CCSD(T) calculation and that of Fedorov et al \cite{fedrov} for the ground state shows a very good concordance with a relative differences of $\Delta < $1.88$\%$ for R$_e$, and $\Delta < $8$\%$ for D$_e$. The spectroscopic constants found using the MP2 approach are also very good agreement in comparison with Pyykk\"{o} \cite{pykk} for the equilibrium distance R$_e$ as well as for the frequency $\omega_e$ ($\Delta$(R$_e$)$<$1.1$\%$, $\Delta$($\omega_e$)$<$6.81$\%$).
 
  New calculations were also carried out using the CCSD(T) and MP2 level of calculations and using the Molpro package for BeK$^+$, and BeRb$^+$. We used a def2QZVPPD+CRENBL ECP10 \cite{hurly} basis set for K and a AQZP \cite{fantin,martins} basis set for Rb. The comparisons between the CIPSI and CCSD(T) calculations show a very good concordance as the relative differences are $\Delta$(R$_e$)$<$3.4$\%$ and $\Delta$($\omega_e$)$<$3.7$\%$ for BeK$^+$ and $\Delta$(R$_e$)$<$5.6$\%$ and $\Delta$($\omega_e$)$<$6.2$\%$ for BeRb$^+$

\begin{figure}
\begin{center}
 \includegraphics[width=0.6\linewidth]{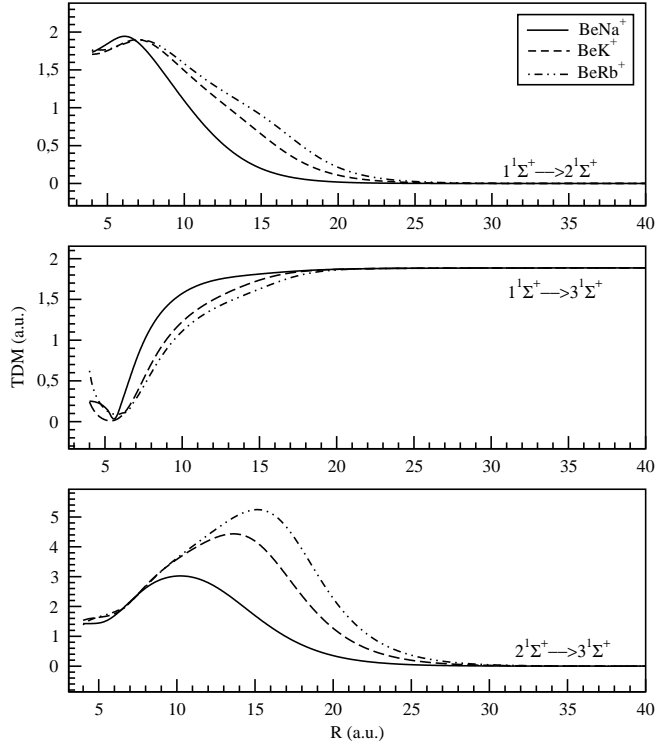}
 \caption {Variation of transition dipole moment as function of internuclear separation between the different electronic states of BeX$^+$ systems.}

 \label{tdp}
 \end{center}
\end{figure}

Furthermore, we have calculated the adiabatic transition dipole
moments from X$^1\Sigma^+$ states to the 2$^1\Sigma^+$ and 3$^1\Sigma^+$ electronic states and between
the first 2$^1\Sigma^+$ and second excited states 3$^1\Sigma^+$ is shown in figure \ref{tdp} as a function internuclear separation. From this plot it is cleared that the transition dipole moments display a very similar shape for all these ionic 
molecular systems. These similar behaviors indicate that the electronic
wave functions behave in same way for all these systems.
Our data,
potential interaction and dipole moments, are also provided as supporting
information for interested scientists to be used for further simulations.

 \section{Part-II: Ion-atom cold collisions}
 \subsection{Results and Discussions}
 In this section we present results on elastic collisions at low energy between alkaline ion (Be$^+$) and alkali atom (X=Na, K, Rb). We consider, ion-atom collision takes place in $2^1\Sigma^+$ molecular potential of BeX$^+$ systems where both atom and ion asymptotically goes to $^2S$ + $^2S$ electronic states. The choice of the initial state is motivated by ongoing experiments \cite{smith} as a single trapped Be$^+$ ion immersed in a Bose-Einstein condensate of alkali atoms X (X=Na, K or Rb) in a magneto-optical trap.

 When the interaction potential $V(r)$ becomes spherically symmetric, the Hamiltonian can be splited into radial part and angular part. Using partial-wave decomposition, the long-range part of the effective potential of ion-atom systems can be expressed as 
\begin{equation}
 V_{eff}=- \frac{1}{2} \left(\frac{C_4}{r^4}+\frac{C_6}{r^6}\right)+\frac{\hbar^2}{2\mu r^2}l(l+1)
\end{equation}
where $r$ is separation between ion and atom, $\mu$ is the reduced mass of the system and $l$ is the partial wave. The term $V_{CB}=\frac{\hbar^2}{2\mu r^2}l(l+1)$ is the centrifugal barrier which is in general higher than kinetic energy of colliding ion-atom pair for $l\neq0$ at all energy. The time independent Schrodinger equation for $l$-th partial wave describing the system as 
\begin{equation}
 \left (-\frac{\hbar^2}{2\mu}\frac{d^2}{dr^2} +V_{eff}(r)\right)\psi_l(kr)=E\psi_l(kr)
 \label{se}
\end{equation}
here $k$ is the wave number related to collisional energy $E=\frac{\hbar^2 k^2}{2\mu}$. The asymptotic form of the wave function is $\psi_l(kr)\sim \sin[kr-l\pi/2 + \eta_l]$ with $\eta_l$ being the phase shift for $l$th partial waves. The above second order partial differential equation (\ref{se}) can be solved by standard Numerov-Cooley  method described elsewhere \cite{numerov}. The total elastic scattering cross section is given by 
\begin{equation}
 \sigma_{el}=\frac{4\pi}{k^2}\sum^\infty_{l=0}\sin^2(\eta_l)
\end{equation}
\begin{figure}
\begin{center}
 \includegraphics[width=0.75\linewidth]{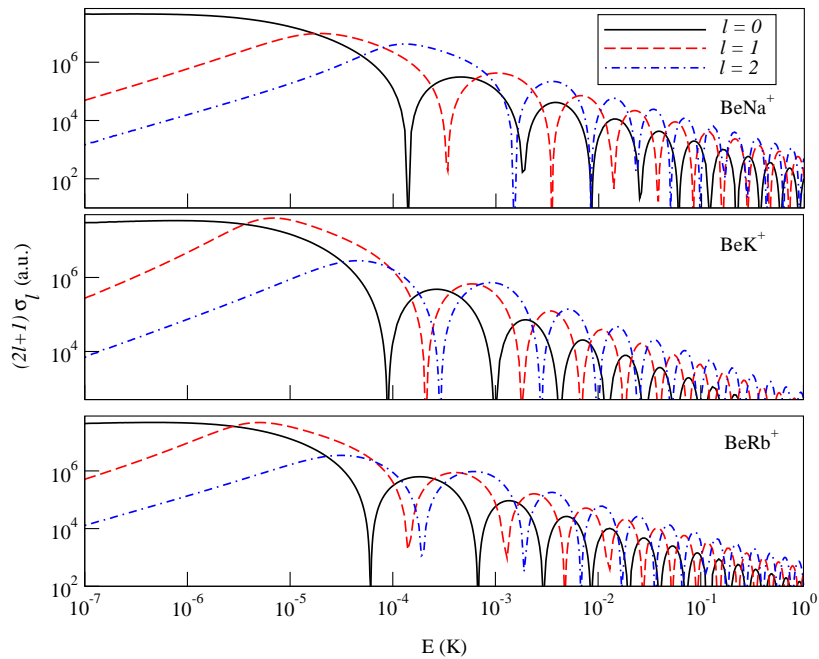}
 \caption {In upper panel partial-wave cross sections are plotted for Be$^+$+Na (2$^1\Sigma^+$) collision as function of collision energy E (in K) for $l=0$ (black solid line), $l=1$ (red dashed line) and $l=2$ (blue double dashed line). Lower two panels show the same for Be$^+$+K and Be$^+$+Rb collision where each system present in 2$^1\Sigma^+$ potential. } 

 \label{partial}
  \end{center}
\end{figure}
 
Firstly, we compute the partial wave elastic scattering cross section for excited 2$^1\Sigma^+$ potential for the BeX$^+$(X=Na, K, Rb) systems. In figure \ref{partial}, we have shown the partial-waves $s(l=0)$, $p(l =1)$ and $d(l=2)$ cross sections against collisional energy E.
We compare our work for the elastic
scattering cross section with the results of hetero-nuclear ion-atom systems \cite{makarov,krych1}. A typical behavior is observed for elastic scattering cross
sections at low energies. This distinctive behavior is a consequence of the
Wigner threshold laws which arise from the analytic properties of
scattering wave functions at low collision energies. From the figure \ref{partial}, it is cleared that the $s$-wave cross section becomes independent of energy when $k\rightarrow0$  while higher order partial waves vary with energy for the three BeX$^+$ systems.

\begin{figure}
\begin{center}
 \includegraphics[width=0.8\linewidth]{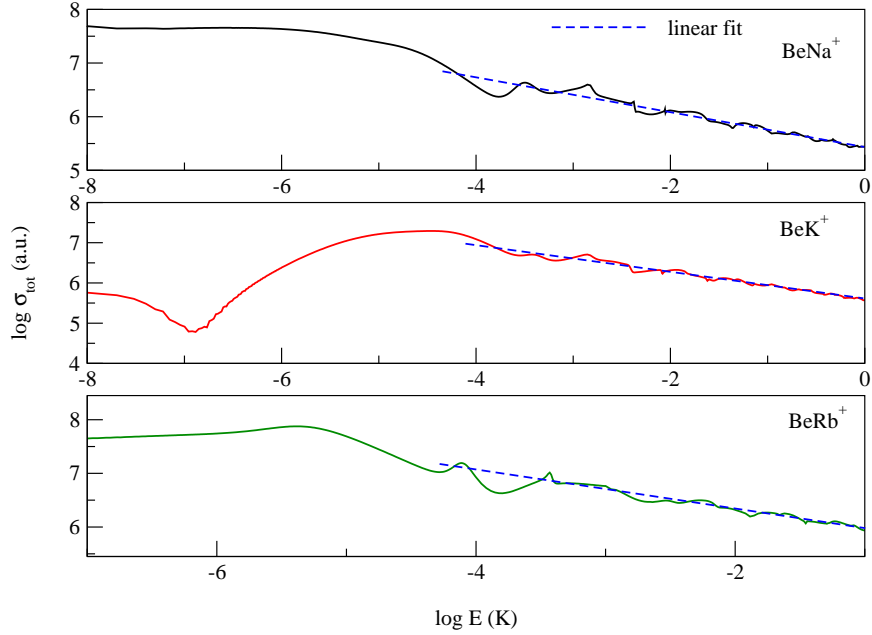}
 \caption {Variation of logarithm of total elastic scattering cross section as function of logarithm energy for Be$^+$+X (X=Na, K, Rb) collision occurring an electronic state 2$^1\Sigma^+$ potential are plotted in three different panels. The dashed line is a linear fit for energies above 10$^{-6}$K. }
 \label{Total}
 \end{center}
\end{figure}
   
As $k\rightarrow0$, the phase
shift for $l$-th partial-wave is $\eta_l\sim k^{2l+1}$
if $l\leqslant (n-3)/2$, otherwise $\eta_l\sim k^{n-2}$
for a long range potential varying as $1/r^n$. An accurate analysis for all these cases
show that the Wigner thershold behavior for these molecular ionic systems
begin to set in as the collision energy decreases below 1$\mu $K. In this
regime, the $s$-wave cross section becomes independent of energy while all
other higher order partial waves cross section goes as $k^2$ . In the case of atom-ion
collisions, the characteristic energy for the occurrence of the $s$-wave
threshold behavior is in the $\mu $K temperature range. But in case of
neutral alkali atom system these behavior is observed for energies around 100$\mu $K mainly due to shorter range of the van der Waals interactions between
neutral atoms. In the Wigner thresold regime, heteronuclear ion-atom elastic collision is the dominant process and charge transfer path is suppressed. The $s$-wave cross section show a minimum at a low energy when the energy increases beyond the Wigner thershold law regime. This may be
related to the Ramsauer-Townsend effect \cite{mitory,reid}.

\begin{figure}
\begin{center}
 \includegraphics[width=0.5\linewidth]{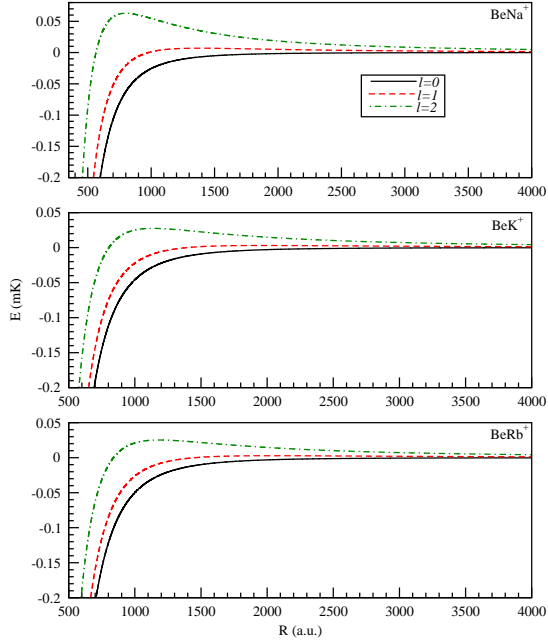}
 \caption {In upper panel, we consider variation of centrifugal energies as a function of internuclear separation for three partial waves $s$-(black solid line), $p$-(red dashed line) and $d$-(green double dashed line) for (BeNa)$^+$ system. In lower two panels show the same results for  (BeK)$^+$ and  (BeRb)$^+$, respectively.}
 \label{Centrifugal}
 \end{center}
\end{figure}

In figure \ref{Total}, we have shown logarithm of total elastic scattering cross section ($\sigma_{tot}$) with respect to logarithm of energy ($E$) for three different ion-atom systems (BeNa)$^+$, (BeK)$^+$ and (BeRb)$^+$ respectively, colliding in the 2$^1\Sigma^+$ electronic state. To obtain the converging result of total scattering cross section in high energy region (greater than 1 mK), we require more than 80 partial waves for both (BeNa)$^+$, (BeK)$^+$  systems and 50 partial waves need for (BeRb)$^+$ system. Since, with the increase in collision energy, more and more number of partial waves start to contribute to $\sigma_{tot}$.   
We note that in all the cases, the $s$-wave
contribution is dominant at energy corresponding to temperature below 10$^{-6}$ K. This can be explained by the effect of the height of the
centrifugal barrier which varies from system to system. Since,
 an interaction potential varies as -$−C_4 /r^4$, the height of the centrifugal barrier
is given by $l(l+1)^{5/2} /(4C^{3/2}_4 \mu^{5/2})$. So, for energies below 1$\mu $K, the effect of the inner part of the potential of the higher partial waves is suppressed by
the centrifugal barrier. In figure \ref{Centrifugal}, we have shown the variation of centrifugal barrier as a function of internuclear separation for our three systems.

At high collision energies, the cross sections are well
represented by power-law functions described as 
\begin{equation}
 \sigma_{tot}\sim\pi \left(\frac{\mu C^2_4}{\hbar^2}\right)\left(1+\frac{\pi^2}{16}\right)E^{-1/3}
 \label{tot}
\end{equation}
Taking logarithm both sides of equation (\ref{tot}), a straight line is obtained obeying the equation of state $\log\sigma_{tot}(E)=-(1/3)E+c_E$ where the slope of the line is -1/3 and the intercept $c_E$ depends on dipole polarizabiliy i.e $C_4$ coefficient of long range-part of the potential. 
From the linear fitting of $\log\sigma_{tot}$ vs $\log E$ of figure \ref{Total}, we have
checked numerically that the slope of the line is quite close to 1/3 for the three systems (BeNa)$^+$, (BeK)$^+$ and (BeRb)$^+$, respectively.

\section{Part-III: Forming molecular ion by STIRAP}

\begin{figure}
\begin{center}
 \includegraphics[width=0.75\linewidth]{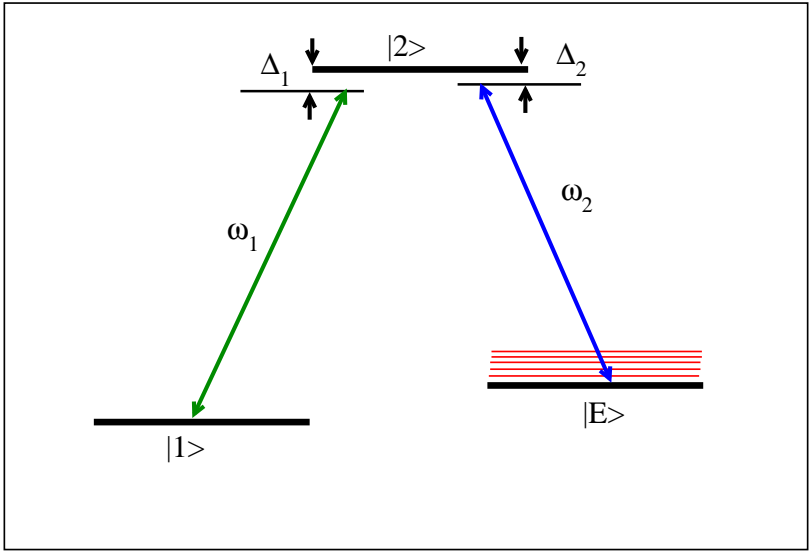}
 \caption {Three states of each BeX$^+$ system looks like a $\Lambda$-type configuration. Here, $\mid E\rangle$ is initially populated ion-atom scattering continuum of 2$^1\Sigma^+$ electronic state, $\mid 2\rangle$ and $\mid 1\rangle$ are the suitable bound states of 3$^1\Sigma^+$ and 2$^1\Sigma^+$ molecular potentials, respectively. The state $\mid 2\rangle$ acts as a intermediate state and $\mid 1\rangle$ becomes final target state. The terms $\omega_1$, $\omega_2$ are the frequencies of the two applied lasers $L_1$, $L_2$  and $\Delta_1$, $\Delta_2$ are the detuning parameters, respectively.}  
 \label{descrete}
 \end{center}
\end{figure}

We consider molecular level structure in $\Lambda$-type configuration involving continuum of states and two bound states of ion-atom systems (BeNa)$^+$, (BeK)$^+$ and (BeRb)$^+$ as shown in figure \ref{descrete}. In this figure, $\mid E\rangle$ represents initially populated ion-atom scattering continuum of first excited molecular potential 2$^1\Sigma^+$ coupled to the ro-vibrational bound state $\mid2\rangle$ of second excited molecular potential 3$^1\Sigma^+$ by laser $L_2$ and the state $\mid2\rangle$ is coupled to ro-vibrational bound state $\mid1\rangle$ of first excited potential by laser $L_1$. Here, $\omega_1$ and $\omega_2$  are the frequency of the two applied lasers $L_1$ and $L_2$, respectively. In this $\Lambda$-type system of BeX$^+$ (X=Na, K, Rb), we want to transfer population from initially populated ion-atom scattering continuum $\mid E\rangle$ to molecular bound state $\mid1\rangle$ via the intermediate state $\mid2\rangle$ by using two lasers $L_1$ and $L_2$ applied in a counter intuitive way. As the molecular ion is formed in the first excited potential, the molecular ion in ground state is likely to be formed by stimulated bound-bound emission or by spontaneous emission.   
 The Hamiltonian for our system under rotating wave approximation is given by  
\begin{equation}
\begin{split}
 H = \hbar(\Delta_1 - \Delta_2)\mid 1\rangle\langle 1\mid + (-\hbar\Delta_2\mid 2\rangle\langle2\mid ) + \int E'\mid E'\rangle\langle E'\mid dE' \\  
 + \int\Lambda_{E'2}\mid E'\rangle\langle2\mid dE'     
      +  \hbar G_1 \mid 1\rangle\langle2\mid  + {\rm H.c.} 
\end{split}
\end{equation} 
where
 $G_1=\frac{1}{\hbar}\langle1\mid {\bf D.{ E}}_{L1}\mid 2\rangle$ is the bound-bound coupling parameter and
$\Lambda_{2E}=\langle2\mid {\bf D.{ E}}_{L2}\mid  E\rangle$ is the free-bound coupling parameter, $\textbf{D}$ being the molecular
electric dipole moment between the two involved levels. The detuning parameters are defined as $\Delta_1 = \omega_1 - (E_{v_2} - E_{v_1})/\hbar  $ 
and $\Delta_2 = \omega_2 - E_{v_2}/\hbar $ where $E_{v_1}$ ($E_{v_2}$) as the energy of the bound state $\mid1\rangle$ ($\mid2\rangle$). The limits of the energy integrals are [$0,\infty$[.
Considering the Fano diagonalization technique, the dressed state of $H$ is given by 
\begin{equation}
\mid E\rangle_{dr}= B_{1E} \mid 1\rangle + B_{2E} \mid 2\rangle + \int F_E(E') \mid E'\rangle dE' 
\end{equation}
where $B_{1E}$, $B_{2E}$, $F_E(E')$ are expansion coefficients. The detailed derivations are done  in the paper \cite{dibyendu}.
As $\mid E\rangle_{dr}$ is energy normalized, so we can write 

\begin{subequations}
\begin{equation}
 \int_{dr} \langle E'\mid E\rangle_{dr} dE=1,
\label{subeqnparta}
\end{equation}
\begin{equation}
\int \mid B_{1E}\mid ^2 dE+\int \mid B_{2E}\mid ^2 dE+\int \int \mid F_E(E'')\mid ^2dE dE''=1,
\label{subeqnpartb}
\end{equation}
\begin{equation}
 P_1+P_2+P_C=1
\end{equation}
\end{subequations}
here $P_1$, $P_2$ and $P_C$ represent the probability that an atom-pair is in the state $\mid 1\rangle$,  $\mid 2\rangle$ and the continuum, respectively. The explicit form of $P_1$ and $P_2$  in terms dimensionless parameter $\beta$ are given by
\begin{equation}
P_1 =\int  \frac{[(\epsilon +\delta_2)^2 +\beta^2]\beta }{(\epsilon +\delta_{12})^2+[(\epsilon +\delta_2)^2+\beta^2]^2 +\beta^2}d\epsilon  
\end{equation}
and
\begin{equation}
P_2=\int  \frac{(\epsilon + \delta_{12})^2[(\epsilon + \delta_2)^2 +\beta^2]\beta }{(\epsilon + \delta_{12})^2[(\epsilon + \delta_2)^2 +\beta^2]^2 + \beta^2}d\epsilon
\end{equation}
where  $\delta_{12}=\hbar(\Delta_2-\Delta_1)/G_1$, $\delta_2=\hbar\Delta_2/G_1$, $\beta=|\Lambda_{2\bar E}| ^2/\hbar G_1$ and $\epsilon=E/G_1$. 

\begin{figure}
\begin{center}
 \includegraphics[width=0.78\linewidth]{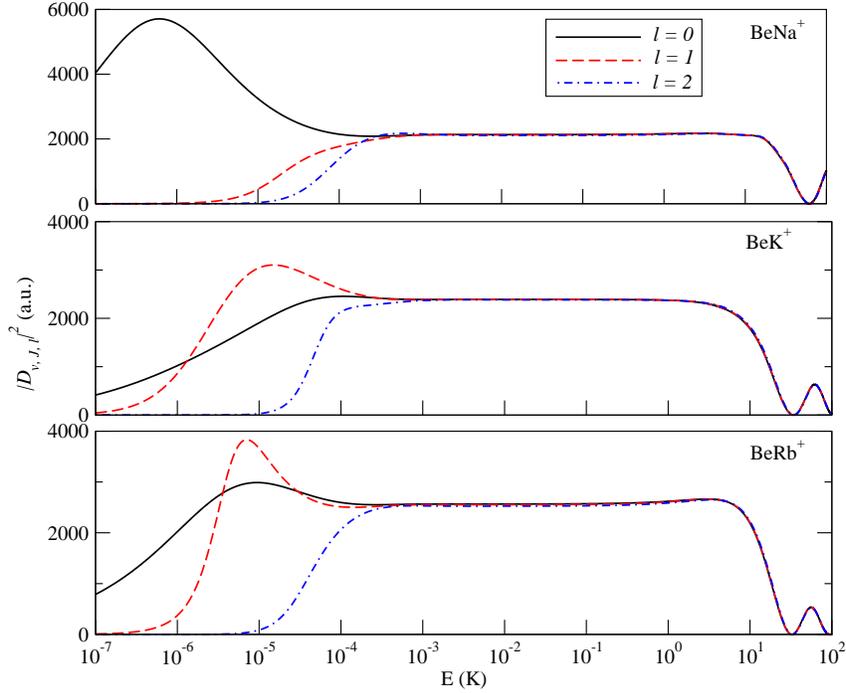}
 \caption {In upper panel, the square of free-bound radial transition dipole moment are plotted with respect to energy E (in K) for continuum of 2$^1\Sigma^+$ electronic state with different partial waves i.e $s$-(solid black line), $p$-(red dashed line) and $d$-(blue double dashed line) and bound vibrational state of 3$^1\Sigma^+$ potential of (BeNa)$^+$ system. Next two panels represent the same for (BeK)$^+$ and (BeRb)$^+$, respectively. }
 \label{FC}
\end{center}
\end{figure}

In case of discrete three level $\Lambda$-type system, the dark-state (DS) condition is achieved only at two-photon resonance condition. But, the systems involving one continuum DS is not obvious. In such systems, the exact two-photon resonance condition is satisfied for a particular energy $\bar E=\hbar (\Delta_1 - \Delta_2)$. For these type systems two-photon resonance condition is changed and it can only be fulfilled when $\Delta_1 > \Delta_2$. Therefore, the bound-bound one-photon detuning $\Delta_1$ must be greater than the one-photon continuum-bound detuning $\Delta_2$ and the DS condition is given by $P_2=\int \mid B_{2E}\mid^2 dE\ll 1$ as this would lead to very small contribution of $\mid2\rangle$ to the dressed state. We referred this state as a quasi-DS. In case of $\Lambda$-like system involving one continuum, the system will remain in quasi-DS when it satisfies the adiabatic condition. The adiabatic condition for this type of system can be expressed as $\left|\frac{d\beta}{dt}\right|\ll k_bT/\hbar$, where $k_b$ is Boltzmann constant and $T$ is temperature of the system.
The detailed of this quasi-DS and adiabatic condition for our model system has been discussed in the reference \cite{dibyendu}. 
\begin{table}[ht]
 \caption{Stimulated linewidth of the selected bound states of BeX$^+$ systems in presence laser $L_1$ having intensity 1 W cm$^{-2}$}
  \centering  
  \begin{tabular}{c c c c }
   \hline\hline
   
     BeX$^+$ system  &      $v_2$      &       $\Gamma(E)$ (KHz)             \\ [0.5ex]
    \hline
         
     (BeNa)$^+$        &      16         &     62.44                  \\
         
     (BeK)$^+$         &      21         &     15.79                \\
         
     (BeRb)$^+$       &      22          &     31.29                   \\ [1ex]

      \hline
         
  \end{tabular}
  \label{sl}
\end{table}

In this section we present result for the formation of molecular ion of the discussed ion-atom colliding systems. To calculate scattering wave functions and preferable bound state wave functions of BeX$^+$ systems, we have to solve partial differential Schrodinger equation (\ref{se}) that we have already discussed in previous section. For determination of transition dipole elements between the two bound states or between the continuum and bound state, we have used calculated transition dipole moment data. Molecular dipole transitions between continuum and bound state or between two ro-vibrational states arises from Franck-Condon principle. According to this principle, for excited vibrational (bound) states, bound–
bound or continuum-bound transitions mainly occur near the turning
points of bound states. In general, highly excited vibrational wave
functions of diatomic molecules or molecular ions have their maximum
amplitude near the outer turning points. Spectral intensity is proportional
to the square of the Franck-Condon overlap integral or Franck-Condon factor (FC). It implies that 
spectral intensity will be significant when the continuum state has a
prominent anti-node near the separation at which outer turning point of
the excited bound state lies. We select ro-vibrational state or bound state ($v_2=16, J=1$) of 3$^1\Sigma^+$ potential for (BeNa)$^+$ system where as (BeK)$^+$ and (BeRb)$^+$ systems, the selected ro-vibrational states are ($v_2=21, J=1$) and ($v_2=22, J=1$), respectively. The choice of this bound state for three different systems depends preferable FC factor  between the scattering wave function of 2$^1\Sigma^+$ potential and ro-vibrational wave function of 3$^1\Sigma^+$ potential, respectively.
The free-bound transition dipole moment element ${\mathcal D}_{v J,\ell}$ can be expressed as 
\begin{equation}      
{\mathcal D}_{v J,\ell} = \langle \phi_{v J}\mid {\mathbf D}(r)\cdot {\mathbf E}_{L_2}\mid \psi_{\ell,k}(r)\rangle = \mid E_{L_2}\mid ^2 \eta_{J,\ell} D_{v, \ell}  \end{equation}
where $\mid \phi_{v J}\rangle$ is the ro-vibrational bound state in 3$^1\Sigma^+$ potential having vibrational and rotational quantum number $v$ and $J$, respectively and $\mid \psi_{\ell,k}(r)\rangle$ is the energy normalized partial wave scattering state of 2$^1\Sigma^+$  potential.  ${\mathbf D}(r)$ is the molecular transition dipole moment between 2$^1\Sigma^+$ and 3$^1\Sigma^+$ potentials of the three BeX$^+$ systems depends on internuclear separation and
$\eta_{J,\ell}  $ is an angular factor where $| \eta_{J,\ell}| \le 1 $. Here, the term $D_{v, \ell}$ poorly depends on rotational quantum number $J$, we can easily neglect $J$-dependence of $D_{v, \ell}$ and it can be expressed as
\begin{equation} 
D_{v, \ell} = \int d r \phi_{v J}(r) D(r) \psi_{\ell,k}(r)
\end{equation}
The stimulated line width is given by $\Gamma(E) = 2 \pi \mid  \eta_{J,\ell} D_{v, \ell}\mid ^2 $. The value of the stimulated line width $\Gamma(E)$ for the selected bound state of BeX$^+$ systems present in 3$^1\Sigma^+$ state are given in table \ref{sl}.

In figure \ref{FC}, we have shown the variation of free-bound radial transition dipole moment $\mid D_{v, \ell}\mid^2$ as a function of energy in different panel for $s$-($l=0$), $p$($l=1$) and $d$-wave($l=2$) scattering state of 2$^1\Sigma^+$ potential and selected $\phi_{v J}$ of 3$^1\Sigma^+$ potential for each BeX$^+$ systems. The value of squre of free-bound radial transition dipole moment at low energy (0.1 $\mu$Kelvin) for (BeNa)$^+$ system present in the upper panel of this figure is greater than the other two systems (BeK)$^+$ and (BeRb)$^+$,  respectively. From figure \ref{FC} it is also cleared that $s$-wave makes finite contribution at very low energy for each of BeX$^+$ systems where as contribution of higher order partial waves are very small. But, in the energy regime above 0.1mK, the contribution of higher partial waves become significant to the dipole transitions.

In our systems quasi-DS formation is possible for a range of collision energies near $\bar E$ that we have discussed earlier and the ion-atom systems will be in quasi-DS for such energy regime. For our three systems, we consider $\bar E=0.1$mKelvin. In order to bring a finite fraction of ion-atom pairs in quasi-DS, we have to set the temperature $k_bT=\bar E$ for a thermal system of ion-atom mixture. Under this condition, if we vary the ratio of continuum-bound and bound-bound coupling parameters $\beta$ subject to the fulfillment of adiabatic condition, then one can achieve STIRAP like process to form cold molecular ions from cold ion-atom pairs.
\begin{figure}
\begin{center}
 \includegraphics[width=0.8\linewidth]{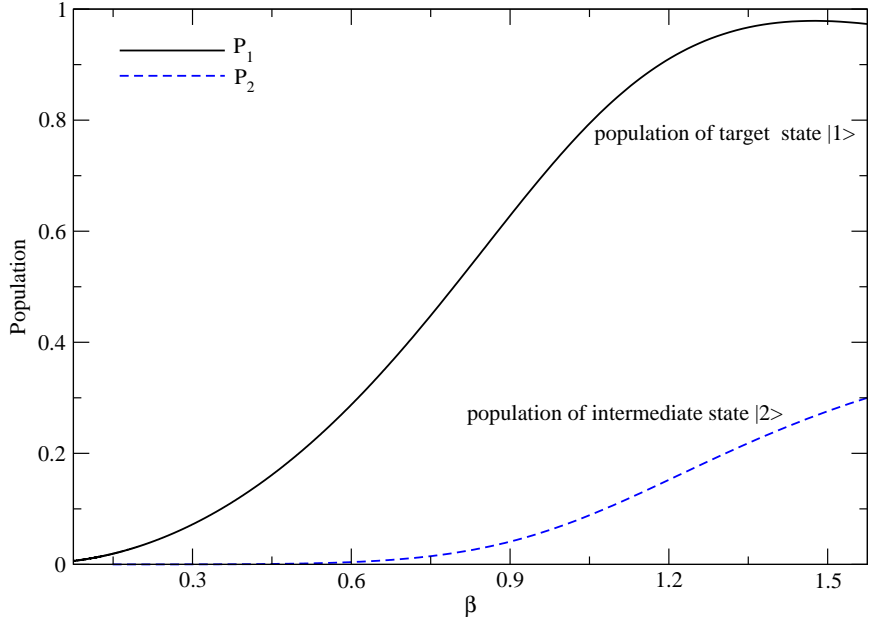}
 \caption {Population of final state $P_1$ (solid black line) and intermediate state $P_2$ (blue dashed line) are plotted as function of $\beta$ for (BeNa)$^+$ system and $\beta=\mid \Lambda_{2\bar E}\mid^2/\hbar G_1$. The nature of the plot remains same both for (BeK)$^+$ and (BeRb)$^+$ systems.}

 \label{Population}
 \end{center}
\end{figure}

In figure \ref{Population}, we have shown the variation of probability of target state $\mid1\rangle$ and intermediate state $\mid2\rangle$ as a function of dimensionless parameter $\beta$ for our three ion-atom colliding systems. The nature of plot for each BeX$^+$ system remains same obtained for different set of parameters. We consider first (BeNa)$^+$ system for which we have taken scaled detuning parameters $\delta_2=0.5$, $\delta_{12}=0.1$ and $\Gamma_{2\bar E}=1.45$ MHz that corresponds to FC factor at an energy 0.1 mK and intensity 1.5 kWcm$^{-2}$ of $L_2$. In case of (BeK)$^+$ and (BeRb)$^+$ systems, detuning parameters remain same but $\Gamma_{2\bar E}$ values are 1.55 MHz and 1.20 MHz with intensities 1.4 kWcm$^{-2}$ and 1.2 kWcm$^{-2}$ of laser $L_2$, respectively. In all three systems, we obtain similar type of plot by varying the parameter $\beta$ from 0.2 to 200. It is clear from figure \ref{Population} that the probability $P_2$ of the intermediate state $\mid2\rangle$ rises significantly with increase in value of $\beta$. Thus, for higher value of $\beta$, the systems come out from quasi-DS regime. But, in the lower range of $\beta$, probability ($P_1$) of target state rises rapidly whereas that of $P_2$ remains close to zero. Therefore, in this regime when bound-bound coupling parameter is much higher compare to the continuum-bound coupling parameter, the systems remain in quasi-DS and population transfer efficiency is near about 55-65 percent for our three BeX$^+$ systems.

\section{Conclusions}
In our study, by calculating ab initio data, potential energy curves have been plotted. Also, the  interaction potentials and spectroscopic constants have been calculated in ground and other two low laying excited states of (BeNa)$^+$, (BeK)$^+$ and (BeRb)$^+$ molecular ionic systems. To the best of our knowledge, for the first time ab initio calculation of the system of alkaline ion (Be$^+$) interacting with alkali atom (K, Rb) is being reported here. However, our data for the BeNa$^+$ ground state were compared with available theoretical data \cite{pykk,fedrov}. The good agreement between our spectroscopic constants and the recent ones of Fedorov et al \cite{fedrov} proves the high accuracy of our calculation and validate the use of the pseudopotential technique. In close connections, we have studied elastic collision physics in a range of energy for these BeX$^+$ (X=Na, K, Rb) systems in which each system follows Wigner thresold law at low energy and the $1/3$ law at higher energy regime. Besides, we explored that by applying STIRAP method molecular ion can be formed in ground electronic potential. It is observed that the system involving bare continuum resembling a $\Lambda$-like configuration, the nature of DS will remain in flickering form resulting in partial population transfer. In our BeX$^+$ systems, we have shown that 55-65 percent population can be transferred by STRAP method for molecular ion formation. We expect that our analysis including ab initio data can serve as a solid foundation for future studies on BeX$^+$ systems.

  {\bf{Acknowledgements}} \\
 Dibyendu Sardar is grateful to CSIR, Government of India, for a financial support. 
This work is also jointly supported by Department of Science and Technology (DST), Ministry of Science and Technology, Govt. of India and Ministry of Higher Education and Scientific Research (MHESR), Govt. of Tunisia, under an India-Tunisia Project for Bilateral Scientific Cooperation.

\end{document}